\documentclass[conference]{IEEEtran}
\IEEEoverridecommandlockouts
\usepackage{cite}
\usepackage{amsmath,amssymb,amsfonts}
\usepackage{algorithmic}
\usepackage{graphicx}
\usepackage{textcomp}
\usepackage{xcolor}
\usepackage{multirow}
\usepackage{numprint}
\npdecimalsign{.}
\newcommand{\eer}[1]{\nprounddigits{2}\numprint{#1}}
\newcommand{\dcf}[1]{\nprounddigits{3}\numprint{#1}}

\def\BibTeX{{\rm B\kern-.05em{\sc i\kern-.025em b}\kern-.08em
    T\kern-.1667em\lower.7ex\hbox{E}\kern-.125emX}}
\begin{document}

\title{Study on the temporal pooling used in deep neural networks for speaker verification}

\author{
\IEEEauthorblockN{Mickael Rouvier}
\IEEEauthorblockA{\textit{LIA - Avignon University}\\
Avignon, France \\
mickael.rouvier@univ-avignon.fr}
\and
\IEEEauthorblockN{Pierre-Michel Bousquet}
\IEEEauthorblockA{\textit{LIA - Avignon University}\\
Avignon, France \\
pierre-michel.bousquet@univ-avignon.fr}
\and
\IEEEauthorblockN{Jarod Duret}
\IEEEauthorblockA{\textit{LIA - Avignon University}\\
Avignon, France \\
jarod.duret@alumni.univ-avignon.fr}
}

\maketitle

\begin{abstract}
The $x$-vector architecture has recently achieved state-of-the-art results on the speaker verification task. This architecture incorporates a central layer, referred to as temporal pooling, which stacks statistical parameters of the acoustic frame distribution.
This work proposes to highlight the significant effect of the temporal pooling content on the training dynamics and task performance.
An evaluation with different pooling layers is conducted, that is, including different statistical measures of central tendency. Notably, 3$\textsuperscript{rd}$ and 4$\textsuperscript{th}$ moment-based statistics (skewness and kurtosis) are also tested to complete the usual mean and standard-deviation parameters.
Our experiments show the influence of the pooling layer content in terms of speaker verification performance, but also for several classification tasks (speaker, channel or text related), and allow to better reveal the presence of external information to the speaker identity depending on the layer content.
\end{abstract}

\section{Introduction}

Speaker recognition refers to the task of verifying the identity claimed by a speaker from that person's voice~\cite{bimbot2004tutorial}. For example, it has been shown useful for speaker diarization~\cite{rouvier2012global}, forensics~\cite{campbell2009forensic} or voice dubbing~\cite{gresse2017acoustic}. 

These last years, Deep Neural Networks (DNN) have allowed to emerge new voice representations, outperforming the state-of-the-art $i$-vector framework~\cite{dehak2010front}. One of this DNN approach seeks to extract an embedding representation of a speaker directly from its acoustic excerpts. This high-level speaker representation is called $x$-vector~\cite{snyder2018x}. The DNN models are trained through a speaker identification task, {\it i.e.} by classifying speech segments into one of $n$ speaker identities. In that context, the different layers of the DNN are trained to extract information for discriminating between different speakers. The idea is to use one of the hidden layer as the speaker representation (the $x$-vector). One of the main advantage is that $x$-vectors produced by the DNN can generalize well to speakers beyond those present in the training set. The benefits of $x$-vectors, in terms of speaker detection accuracy, have been demonstrated during the recent evaluation campaigns: NIST SRE~\cite{villalba2018jhu,lee2019nec,rouvier2019}, VoxCeleb 2020~\cite{thienpondt2020idlab,brummerbut+,torgashov2020id}, SdSVC~\cite{villalba_jhu_2019,bousquet_sdsvc_2019,barbadillo_2019}...

In the $x$-vector framework, the DNN uses a stack of convolution layers followed by a temporal pooling layer that computes the mean and standard deviation of an input sequence, in order to filter and capture the speaker characteristics throughout the recording. The temporal pooling is a very critical part of the DNN as it compacts the information along the full recording into a single vector representation. One of the main goals of temporal pooling is to capture only the salients part of the utterance in a compact representations, while removing irrelevant details. For this reason, the selection of a good temporal pooling in the model is important since it has a significant effect on the task performance.

Moreover, the authors in~\cite{9003979,wang2017does} found that, in addition to the speaker-related information, the extracted $x$-vector contains meta-information such as session, speaking style, lexical content... The hypothesis in our study is that the types and proportions of such meta-information captured by the $x$-vector are greatly depending on the statistical parameters picked up to make up the pooling layer. As a consequence, well combining distinct DNN architectures, in terms of pooling content, could play a significant role in filtering out such unwanted information and, thus, better focusing the system on the goal of speaker discrimination.

In this study, we propose to evaluate the performance of various pooling contents, as well as their combination. In addition to traditional pooling, we propose to evaluate two new poolings : \textit{skewness-pooling} and \textit{kurtosis-pooling}. To further validate our hypothesis about pooling and information filtering, we experimentally evaluate information contained inside the $x$-vectors depending on various pooling contents, through numerous applications: speaker gender, speaker nationality, augmentation type, words recognition....

The papers is organized as follows: Section~\ref{sec:resnet} summarizes the $x$-vector approach.  Section~\ref{sec:pooling} defines the different poolings used in our study. Section~\ref{sec:classification} presents the classifiers and probing tasks, and classifiers for the probing tasks. In Section~\ref{sec:expe_results}, we analyze the results of the probing tasks and present results for our new $x$-vector based system on speaker verification. A conclusion is finally provided in Section~\ref{sec:conclusion}.

\section{$x$-vector based on ResNet}
\label{sec:resnet}

An $x$-vector is a high-level speaker features extracted from DNN models trained through a speaker identification task. The $x$-vector extractor proposed in this paper is a variant based on ResNet~\cite{zeinali2019but}. The detailed topology of the used ResNet is shown in Table~\ref{tbl:resnet34}. The DNN model for extracting $x$-vectors consists of three modules: a \emph{frame-level} feature extractor, a \emph{statistics-level} layer, and \emph{segment-level} representation layers. 

\begin{itemize}
\item The \emph{frame-level} component is based on the well-known ResNet34 topology. The component is composed of four residual blocks. This network uses 2-dimensional features as input and process them using 2-dimensional Convolutional Neural Networks (CNN) layers.

\item The \emph{statistics-level} component is an essential component that converts from a variable length speech signal into a single fixed-dimensional vector. The statistics-level is composed of one layer: the statistics-pooling, which aggregates over frame-level output vectors of the DNN and computes their mean and standard deviation.

\item The \emph{segment-level} component maps the segment-level vector to speaker identities. The mean and standard deviation are concatenated together and forward to additional hidden layers and finally to softmax output layer.
\end{itemize}

\begin{figure*}[h]
\center
\includegraphics[height=110px]{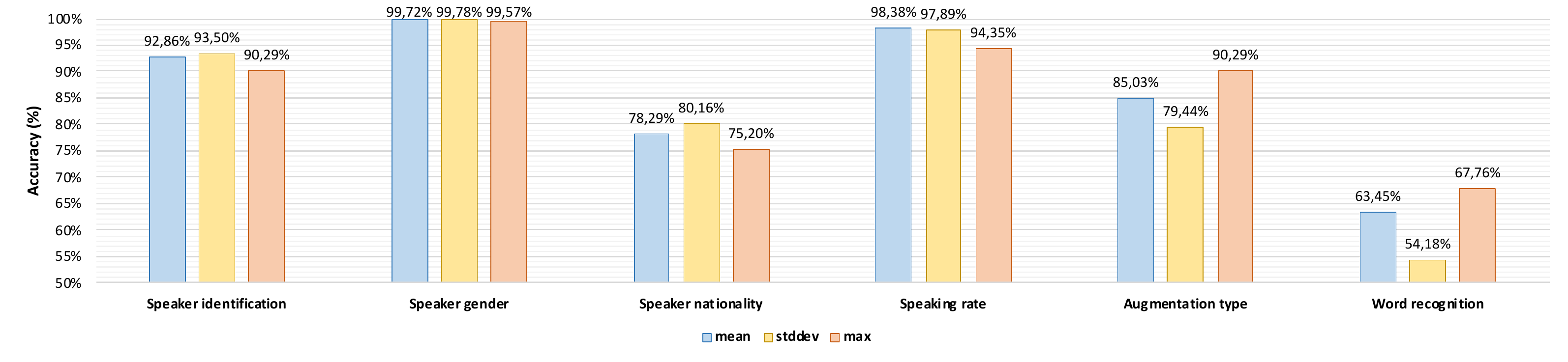}
\caption{Results obtained by using different pooling layer on different classification tasks.}
\label{fig:classification_task}
\end{figure*}

\begin{table}[t]
    \label{tbl:resnet34}
    \caption{The proposed ResNet34 architecture. $N$ in the last row is the number of speakers. Batch-norm and ReLU layers are not shown. The dimensions are (FrequencyxChannels×Time). The input comprises 60 filter bank from speech segments. During training we use a fixed segment length of 400.}
    \begin{tabular}{l c c}
        \hline
        \hline
        \textbf{Layer name}   & \textbf{Structure}          & \textbf{Output} \\
        \hline
        Input                 & --                          & 60 $\times$ 400 $\times$ 1  \\
        Conv2D-1              & 3 $\times$ 3, Stride 1      & 60 $\times$ 400 $\times$ 128 \\
        \hline
        ResNetBlock-1         & $\begin{bmatrix} 3 \times 3, 128  \\ 3 \times 3, 128  \end{bmatrix} \times 3$  , Stride 1& $60\times 400 \times128$  \\
        ResNetBlock-2         & $\begin{bmatrix} 3 \times 3, 128  \\ 3 \times 3, 128  \end{bmatrix} \times 4$, Stride $2$ & $30 \times 200 \times 128$  \\
        ResNetBlock-3         & $\begin{bmatrix} 3 \times 3, 128 \\ 3 \times 3, 256 \end{bmatrix} \times 6$, Stride $2$ & $15 \times 100  \times 256$ \\
        ResNetBlock-4         & $\begin{bmatrix} 3 \times 3, 256 \\ 3 \times 3, 256 \end{bmatrix} \times 3$, Stride $2$ & $8  \times 50  \times 256$ \\
        \hline
        Pooling          & --                & $8 \times 256$                 \\
        Flatten               & --                & $2048$                            \\
        \hline
        Dense1                & --                & $256$                            \\
        Dense2 (Softmax)      & --                & $N$                              \\
        \hline
        Total                 & --                & --                             \\
        \hline
        \hline
    \end{tabular}
\end{table}

The DNN is trained using ArcFace softmax to classify speakers contained in the training set. The ResNet uses 60-dimensional Filter bank features as input, extracted from 25ms audio signal, mean-normalized over a sliding window of up to 3 seconds. Unvoiced frames are filtered out from the utterances using a Voice Activity Detection (VAD) based on signal energy.

In order to increase the diversity of the acoustic conditions in the training set, a data augmentation strategy is used, which adds four corrupted copies of the original recordings to the training list. The recordings are corrupted by adding noise, music and mixed speech (babble) drawn from the MUSAN database~\cite{snyder2015musan} and adding reverberation by using simulated Room Impulse Responses (RIR).

\section{Studied pooling strategies}
\label{sec:pooling}

The main goal of the pooling operation is to aggregate all the outputs given in the \textit{frame-level} into a compact vector. The most commonly used pooling strategies are : \textit{max-pooling}, \textit{mean-pooling} and \textit{standard-deviation-pooling}. We propose to evaluate two new pooling strategies : \textit{skewness-pooling} and \textit{kurtosis-pooling}.

\begin{itemize}
    \item \textbf{max-pooling} : The max function is the most common choice for the pooling. This operation aggregates all vectors present in frame-level component and calculates the maximum, or largest, value. The max pooling is calculated as follows:
    
    \begin{equation}
        max = \max_{i=1}^{n} x_{i}
    \end{equation}

    \item \textbf{mean-pooling} : The mean computes the average of each vector present at the frame-level. The mean pooling is calculated as follows:
    
    \begin{equation}
        \mu = \frac{1}{n} \sum_{i=1}^{n} x_{i}
    \end{equation}

    \item \textbf{standard-deviation-pooling} : The standard-deviation is a measure of variance ({\it i.e.} dispersion) of a series. The standard-deviation pooling is calculated as follows:
    
    \begin{equation}
        \sigma = \sqrt{ \frac{1}{n} \sum_{i=1}^{n} ( x_{i} - \mu )^{2} }
    \end{equation}

    \item \textbf{skewness-pooling} : Skewness provides information about the symmetry of a distribution around the mean. In the case of unimodal distribution, negative skew indicates that the main tail is on the left side of the distribution and positive skew indicates that it is on the right side. The skewness pooling is calculated as follows:
    
    \begin{equation}
        Skew = \frac{1}{n} \sum_{i=1}^{n} ( \frac{ x_{i} - \mu}{\sigma} )^{3}
    \end{equation}

    \item \textbf{kurtosis-pooling} : the kurtosis is a measure of flattening, providing information about how fat/heavy are the tails of a distribution (and hence, how peaked/flat around its mode), therefore about how frequent extreme deviations (or outliers) are from the average value. The kurtosis pooling is calculated as follows:
    
    \begin{equation}
        Kurt = \frac{1}{n} \sum_{i=1}^{n} ( \frac{ x_{i} - \mu}{\sigma} )^{4}
    \end{equation}

\end{itemize}

It is worth noting that there are two kinds of pooling. The max-pooling provides information on features (if the feature is present) and is a non-parametric measure, unlike the mean, standard-deviation, skewness and kurtosis poolings, which provide information on the statistical distribution. The latter are respectively coming from the first, second, third and fourth order moment, and are belonging to the set of values referred to as ``measures of central tendency'' in the field of statistics.

\section{Classification tasks on statistical pooling}
\label{sec:classification}
To assess the assumption proposed in the introduction, we have to reveal the link between the statistical parameters selected to make up the $x$-vector and the patterns ({\it i.e.} the meta-information cited above) contained in the resulting $x$-vector.

To do that, the following classification tasks are carried out, each time with varying pooling configurations :

\begin{itemize}
    \item \textbf{Speaker identification task} : this task measures to what extent the $x$-vectors encode the speaker’s identity, which is crucial for the speaker recognition task. The evaluation set contains 106 different speakers and we report the recognition accuracy.
    \item \textbf{Speaker gender task} : this task measures whether the $x$-vectors can distinguish between gender ({\it i.e.} male or female). We train a two-classes classifier and report the classification accuracy.
    \item \textbf{Speaker nationality task} : this task measures whether the $x$-vectors can distinguish between speaker's nationality. The evaluation set contains 35 different nationalities and we report the classification accuracy.
    \item \textbf{Speaking rate task} : we augment all utterances by 3-way speed perturbation with rates of 0.9, 1.0 and 1.1. This task measures whether the $x$-vectors can capture information on speaking rate. We train a three-classes classifier and report the accuracy of recognition.
    \item \textbf{Augmentation type task} : this task measures whether the $x$-vectors can distinguish the type of data augmentation : noise, music, speech or no augmentation. We train a four-class classifier and provide the accuracy.
    \item \textbf{Word recognition task} : this task measures whether the $x$-vector can capture information about words in the utterance. We select the 25 most-frequent words and set up a classifier that predicts, for each word, whether the word is present or not. The average accuracy of correctly identified words is reported.
\end{itemize}

The ability of the $x$-vector to discriminate between these various tasks, among the different pooling strategies, will provide us information about the filtering-out property of the statistical measures, when used as components of a DNN-pooling layer. In other words, if a pattern of these tasks is present in the $x$-vector, we can train a classifier to recognize it and the performance of the classifier should depend on how well the pattern is embedded in the speaker representation.
Let us note that the first four tasks presented above are about speaker-related information (speaker identification, speaker gender, speaker nationality, speaking rate) while the last two are about text- and channel-related information.

Our hypothesis is that some pooling operations could more easily provide information about features (if the feature is present). Even if the DNN models are trained for the speaker identification task, these kind of poolings would focus more on the biases present in the corpora in order to more easily identify the speakers ({\it i.e.} speakers who always speak in the same environment or microphone). On the other hand, some pooling strategies, due to their structure, would have more difficulties in indicating the presence or absence of a parameter and therefore would focus more on the speaker's parameters.

\section{Experiments and results}
\label{sec:expe_results}

This section describes the experimental setup in terms of dataset and evaluation protocol.

\subsection{Experimental Protocol}

Concerning the speaker verification task, the $x$-vector extractors are trained on the VoxCeleb2 dataset~\cite{chung2018voxceleb2}, only on the development partition, which contains speech excepts from 5,994 speakers with a 16 Khz sampling rate. The trained $x$-vectors are assessed on the Speakers in the Wild (SITW) core-core task~\cite{mclaren2016speakers} and Voxceleb1-E Cleaned~\cite{nagrani2017voxceleb} dataset with a 16 KHz sampling rate. Note that the development set of VoxCeleb2 is completely disjoint from the VoxCeleb1 dataset ({\it i.e.} no speaker in common).

Concerning the classification task, our objective is to evaluate the $x$-vector obtained by varying pooling configurations on the tasks presented in the previous section. As in~\cite{9003979,wang2017does}, for each classification task, we use a   MultiLayer Perceptron (MLP) classifier with a single hidden layer and ReLU activations. The hidden layer size is fixed at 500 for all the different tasks. We used the Librispeech~\cite{panayotov2015librispeech} for word recognition task and Voxceleb1 for all others tasks. For all the tasks we trained on 80\% of this data and evaluated on the remaining 20\%.

\subsection{Performance criterion used in speaker verification}

Equal Error Rate (EER) and Detection Cost Function (DCF) are used as the performance criterion of speaker verification. EER is the threshold value such that false acceptance rate and miss rate are equals.

\subsection{Classification task}
Figure~\ref{fig:classification_task} reports results of the classification tasks described in Section \ref{sec:classification}.
It can be observed that the $x$-vectors extracted from models using mean and standard-deviation as pooling layer achieve the best performance on the tasks related to speaker : speaker identification, speaker gender, speaker nationality and speaking rate. However, the $x$-vectors extracted from model that use max-pooling achieve better accuracy on the tasks : augmentation type and word recognition.
These experiments highlight the importance of the choice of the pooling layer in the architecture.
They show that the $x$-vector, which is fitted to focus on the speaker verification task, actually embeds different biases of the corpus. Moreover, it is also shown that these biases are strongly depending on the statistical parameters chosen to make up the pooling layer.

\subsection{Speaker verification}

Table~\ref{tbl:single-pooling} shows results obtained with the $x$-vectors-based systems using a single statistical measure for pooling. The system that obtains the best results is the one using the standard-deviation pooling (denoted as \emph{std}). This results is very surprising since it means that the automatic speaker verification is not done on speaker-specific traits but on their variations. It can be observed that systems using mean (denoted as \emph{mean}) and standard-deviation pooling better perform than systems using the max pooling (system called \emph{max}). This results tends to show that in the context of speaker verification task, the pooling layer must treat the embeddings given by the frame-level component as a statistical distribution and not as non-parametric measure.

Lastly, the systems using skewness (denoted as \emph{skew}) or kurtosis (denoted as \emph{kurto}) pooling yield very bad results. It can be explained by the fact that the skewness and kurtosis poolings do not provide as much information as other poolings about the speaker signal of voice.
\begin{table}[h]
\center
\caption{Results obtained by systems using a single pooling. The systems called : max, mean, std, skew and kurto refer respectively to maximum, mean, standard-deviation, skewness and kurtosis poolings.}
\begin{tabular}[|c]{|l|c|c|c|c|c|c|}
\hline
\textbf{System} & \multicolumn{2}{c|}{\textbf{VoxCeleb1}} & \multicolumn{2}{c|}{\textbf{VoxCeleb1}} & \multicolumn{2}{c|}{\textbf{SITW}}\\
& \multicolumn{2}{c|}{\textit{-E cleaned}} & \multicolumn{2}{c|}{\textit{-H cleaned}} & \multicolumn{2}{c|}{\textit{core-core}}\\
 & EER & DCF & EER & DCF & EER & DCF \\\hline
max & \eer{1.501} & \dcf{0.1546} & \eer{2.539} & \dcf{0.2313} & \eer{1.777} & \dcf{0.1465} \\
mean  & \eer{1.445} & \dcf{0.1609} &  \eer{2.45} & \dcf{0.2357} & \eer{1.722} &  \dcf{0.1508}  \\
std  & \textbf{\eer{1.29}} & \textbf{\dcf{0.1357}} &  \textbf{\eer{2.154}} & \textbf{\dcf{0.2042}} & \textbf{\eer{1.394}} &  \textbf{\dcf{0.1377}}  \\
skew  & \eer{46.19} & \dcf{0.9938} &  \eer{46.33} & \dcf{0.9954} & \eer{31.44} &  \dcf{0.9510}  \\
kurto  & \eer{49.57} & \dcf{0.9938} &  \eer{49.81} & \dcf{0.9954} & \eer{39.18} &  \dcf{0.9510}  \\
\hline
\end{tabular}
\label{tbl:single-pooling}
\end{table}

Table~\ref{tbl:multiple-pooling} shows results obtained by systems using multi-statistical poolings. This operation is done by concatenating, for each system, the outputs of different statistical poolings into a unique layer. Let us note that all the different combinations of poolings have been tested, but only the most interesting results are reported. The best performance is obtained by the system concatenating mean, standard-deviation and skewness (denoted as \emph{mean-std-skew)}. The skewness pooling achieves a very slight improvement compared to the baseline system, which uses the mean and standard-deviation pooling (denoted as \emph{mean-std}). The information about the distribution tails (kurtosis pooling) did not improve performance. Also, it can be observed that, in the context of speaker verification, it is better to use a pooling that brings information on statistical measures of central tendency rather than on non-parametric values (the maximum).

\begin{table}[h]
\center
\caption{Results obtained by systems using multi-statistical poolings.}
\begin{tabular}[|c]{|l|l|c|c|c|c|c|c|}
\hline
\textbf{System}  & \multicolumn{2}{c|}{\textbf{VoxCeleb1}} & \multicolumn{2}{c|}{\textbf{VoxCeleb1}} & \multicolumn{2}{c|}{\textbf{SITW}}\\
& \multicolumn{2}{c|}{\textit{-E cleaned}} & \multicolumn{2}{c|}{\textit{-H cleaned}} & \multicolumn{2}{c|}{\textit{core-core}}\\
 & EER & DCF & EER & DCF & EER & DCF \\\hline
max-mean  & \eer{1.453} & \dcf{0.1510} &  \eer{2.461} & \dcf{0.2255} & \eer{2.078} &  \dcf{0.1877}  \\
max-std  & \eer{1.418} & \dcf{0.1517} &  \eer{2.358} & \dcf{0.2225} & \eer{1.968} &  \dcf{0.1631}  \\
\hline
mean-skew  & \eer{1.331} & \dcf{0.1481} &  \eer{2.268} & \dcf{0.2165} & \eer{1.722} &  \dcf{0.1853}  \\
std-skew  & \eer{1.283} & \dcf{0.1387} &  \eer{2.182} & \dcf{0.2027} & \eer{1.449} &  \textbf{\dcf{0.1329}}  \\
mean-std  & \eer{1.252} & \dcf{0.1414} &  \eer{2.113} & \dcf{0.1997} & \eer{1.422} &  \dcf{0.1346}  \\
mean-std-skew & \textbf{\eer{1.239}} & \textbf{\dcf{0.1366}} & \textbf{\eer{2.111}} & \textbf{\dcf{0.1984}} & \textbf{\eer{1.394}} & \dcf{0.1383} \\
\hline
\end{tabular}
\label{tbl:multiple-pooling}
\end{table}

\begin{table*}[h]
\center
\caption{Results obtained by merging systems made up of different pooling statistics. The terms mean, std and skew refer respectively to the mean, standard-deviation and skewness statistics used to fill the pooling layer of the DNN. The $\oplus$ sign indicates the fusion of scores}.
\begin{tabular}[|c]{|l|c|c|c|c|c|c|}
\hline
\textbf{System}  & \multicolumn{2}{c|}{\textbf{VoxCeleb1}} & \multicolumn{2}{c|}{\textbf{VoxCeleb1}} & \multicolumn{2}{c|}{\textbf{SITW}}\\
& \multicolumn{2}{c|}{\textit{-E cleaned}} & \multicolumn{2}{c|}{\textit{-H cleaned}} & \multicolumn{2}{c|}{\textit{core-core}}\\
 & EER & DCF & EER & DCF & EER & DCF \\\hline
(mean)$\oplus$(mean-skew)  & \eer{1.249} & \dcf{0.1443} &  \eer{2.175} & \dcf{0.2094} & \eer{1.449} &  \dcf{0.1421}  \\
(std)$\oplus$(std-skew)  & \eer{1.175} & \dcf{0.1298} &  \eer{2.006} & \dcf{0.1913} & \eer{1.394} &  \dcf{0.1311}  \\
(mean-std)$\oplus$(mean-std-skew)  & \textbf{\eer{1.153}} & \dcf{0.1309} &  \textbf{\eer{1.974}} & \textbf{\dcf{0.1902}} & \textbf{\eer{1.312}} &  \textbf{\dcf{0.1288}}  \\
\hline
(mean)$\oplus$(std)  & \eer{1.214} & \dcf{0.1362} &  \eer{2.105} & \dcf{0.2031} & \eer{1.449} &  \dcf{0.1367}  \\
(mean)$\oplus$(mean-std)  & \eer{1.238} & \dcf{0.14} &  \eer{2.114} & \dcf{0.2053} & \eer{1.449} &  \dcf{0.139}  \\
(std)$\oplus$(mean-std)  & \eer{1.162} & \dcf{0.1299} &  \eer{1.965} & \dcf{0.1923} & \eer{1.422} &  \dcf{0.1317}  \\
\hline
(mean-skew)$\oplus$(std-skew)  & \eer{1.149} & \dcf{0.1317} &  \eer{2.003} & \dcf{0.1913} & \eer{1.367} &  \dcf{0.134}  \\
(mean-skew)$\oplus$(mean-std-skew)  & \eer{1.161} & \dcf{0.1304} &  \eer{2.001} & \dcf{0.1915} & \eer{1.285} &  \dcf{0.1361}  \\
(std-skew)$\oplus$(mean-std-skew)  & \eer{1.149} & \textbf{\dcf{0.1283}} &  \eer{1.983} & \dcf{0.1872} & \eer{1.394} &  \dcf{0.1309}  \\\hline
\end{tabular}
\label{tb:fusing-pooling-1}
\end{table*}

Table~\ref{tb:fusing-pooling-1} summarizes results of fusion of scores carried out on various pooling-content based systems. The fusion of systems is done at the score level, by simply averaging the scores provided by the systems with equal weights. The goal of these experiments is to assess the contribution of the skewness measure to a fusing approach. Let us note that, for all the systems tested, the mini-batch and the weight initialization of the DNN are the same.

In the upper part of Table~\ref{tb:fusing-pooling-1} (rows 1 to 3) we propose to fuse the scores of an initial system to those of its version with the additional skewness statistic. Each time, the fusion leads to a significant gain of performance. Moreover, the fusion reported in row 3 of the Table significantly improves performance compared to the previous single best one of Table~\ref{tbl:multiple-pooling}, by a relative gain of around 7\%.

The lower part of the table compares the fusions of different systems without (rows 4 to 6) or with (rows 7 to 9) the skewness statistic added to the pooling. The reported results show that fusion of systems using several pooling-layer compositions and, also, inclusion of the skewness statistic, contributes to a significant enhancement of performance.

\section{Conclusion}
\label{sec:conclusion}
Extracting speaker embeddings for speaker recognition by using deep neural network approaches has achieved remarkable results in recent years, when compared to traditional GMM-based probabilistic supervector or i-vector frameworks. It has been noticed that the resulting fixed-size representation of an utterance (referred to as $x$-vector) embeds, in addition to speaker-related information, meta-information such as session, speaking style or lexical content.
In this study, we show that the parts of desired information (the one used to discriminate between speakers) and of additional meta-information captured into the utterance representation are significantly dependent on the parameters of the frame distribution that are selected to make up the DNN pooling layer. This central layer, between the acoustic parameters and the speaker identity, usually stacks the 1$\textsuperscript{st}$ and centered-2$\textsuperscript{nd}$ order statistics of the frames (mean and standard deviation). Here, the normalized 3$\textsuperscript{rd}$ and 4$\textsuperscript{th}$ moments (respectively skewness, which measures the asymmetry of the probability distribution, and kurtosis, for the flattening level) are also foreseen and implemented, as well as the non-parametric maximum value. Moreover, several combinations of these five measures are studied. Experiments carried out for our analysis show non-negligible relations between these measures, when used as pooling layer components, and the ability of the resulting $x$-vectors to detect some meta-information such as gender, nationality, speaking rate, augmentation type and word recognition.

On the other hand, the fusion of speaker recognition systems based on various DNN architectures has proven to be beneficial in terms of speaker detection accuracy. This study shows that, given an architecture (here ResNet with angular margin), varying the only content of the statistic-level components provides a set of subsystems which are able, by a simple fusion (i.e. with equal weights in our experiments, to avoid weak conclusions in terms of robustness), to greatly improve performance of the speaker detection task.

\section*{\textbf{Acknowledgement}}
\label{s:acknowledge}

This research was supported by the ANR agency (Agence Nationale de la Recherche), RoboVox project (ANR-18-CE33-0014).

\bibliographystyle{IEEEtran}
\bibliography{mybib}

% Generated by IEEEtran.bst, version: 1.13 (2008/09/30)
\begin{thebibliography}{10}
\providecommand{\url}[1]{#1}
\csname url@samestyle\endcsname
\providecommand{\newblock}{\relax}
\providecommand{\bibinfo}[2]{#2}
\providecommand{\BIBentrySTDinterwordspacing}{\spaceskip=0pt\relax}
\providecommand{\BIBentryALTinterwordstretchfactor}{4}
\providecommand{\BIBentryALTinterwordspacing}{\spaceskip=\fontdimen2\font plus
\BIBentryALTinterwordstretchfactor\fontdimen3\font minus
  \fontdimen4\font\relax}
\providecommand{\BIBforeignlanguage}[2]{{%
\expandafter\ifx\csname l@#1\endcsname\relax
\typeout{** WARNING: IEEEtran.bst: No hyphenation pattern has been}%
\typeout{** loaded for the language `#1'. Using the pattern for}%
\typeout{** the default language instead.}%
\else
\language=\csname l@#1\endcsname
\fi
#2}}
\providecommand{\BIBdecl}{\relax}
\BIBdecl

\bibitem{bimbot2004tutorial}
F.~Bimbot, J.-F. Bonastre, C.~Fredouille, G.~Gravier, I.~Magrin-Chagnolleau,
  S.~Meignier, T.~Merlin, J.~Ortega-Garc{\'\i}a, D.~Petrovska-Delacr{\'e}taz,
  and D.~A. Reynolds, ``A tutorial on text-independent speaker verification,''
  \emph{EURASIP Journal on Advances in Signal Processing}, vol. 2004, no.~4, p.
  101962, 2004.

\bibitem{rouvier2012global}
M.~Rouvier and S.~Meignier, ``A global optimization framework for speaker
  diarization,'' in \emph{IEEE Odyssey - The Speaker and Language Recognition
  Workshop}, 2012.

\bibitem{campbell2009forensic}
J.~P. Campbell, W.~Shen, W.~M. Campbell, R.~Schwartz, J.-F. Bonastre, and
  D.~Matrouf, ``Forensic speaker recognition,'' \emph{IEEE Signal Processing
  Magazine}, vol.~26, no.~2, pp. 95--103, 2009.

\bibitem{gresse2017acoustic}
A.~Gresse, M.~Rouvier, R.~Dufour, V.~Labatut, and J.-F. Bonastre, ``Acoustic
  pairing of original and dubbed voices in the context of video game
  localization,'' in \emph{Interspeech}, 2017.

\bibitem{dehak2010front}
N.~Dehak, P.~J. Kenny, R.~Dehak, P.~Dumouchel, and P.~Ouellet, ``Front-end
  factor analysis for speaker verification,'' \emph{IEEE Transactions on Audio,
  Speech, and Language Processing (TASLP)}, vol.~19, no.~4, pp. 788--798, 2010.

\bibitem{snyder2018x}
D.~Snyder, D.~Garcia-Romero, G.~Sell, D.~Povey, and S.~Khudanpur, ``X-vectors:
  Robust dnn embeddings for speaker recognition,'' in \emph{IEEE International
  Conference on Acoustics, Speech and Signal Processing (ICASSP)}.\hskip 1em
  plus 0.5em minus 0.4em\relax IEEE, 2018, pp. 5329--5333.

\bibitem{villalba2018jhu}
J.~Villalba, N.~Chen, D.~Snyder, D.~Garcia-Romero, A.~McCree, G.~Sell,
  J.~Borgstrom, F.~Richardson, S.~Shon, F.~Grondin \emph{et~al.}, ``The jhu-mit
  system description for nist sre18,'' 2018.

\bibitem{lee2019nec}
K.~A. Lee, H.~Yamamoto, K.~Okabe, Q.~Wang, L.~Guo, T.~Koshinaka, J.~Zhang, and
  K.~Shinoda, ``The nec-tt 2018 speaker verification system.'' in
  \emph{Interspeech}, 2019, pp. 4355--4359.

\bibitem{rouvier2019}
P.-M.~B. Mickael~Rouvier, ``The lia system description for nist sre 2019,''
  2019.

\bibitem{thienpondt2020idlab}
J.~Thienpondt, B.~Desplanques, and K.~Demuynck, ``The idlab voxceleb speaker
  recognition challenge 2020 system description,'' 2020.

\bibitem{brummerbut+}
N.~Brummer, L.~Burget, O.~Glembek, P.~Matejka, L.~Mo{\v{s}}ner, O.~Novotn{\`y},
  O.~Plchot, J.~Rohdin, A.~Silnova, T.~Stafylakis \emph{et~al.}, ``But+ omilia
  system description voxceleb speaker recognition challenge 2020.''

\bibitem{torgashov2020id}
N.~Torgashov, ``Id r\&d system description to voxceleb speaker recognition
  challenge 2020,'' 2020.

\bibitem{villalba_jhu_2019}
J.~Villalba and N.~Dehak, ``The jhu system description for sdsv2020
  challenge,'' 2019.

\bibitem{bousquet_sdsvc_2019}
M.~R. Pierre-Michel~Bousquet, ``The lia system description for sdsv challenge
  task 2,'' 2019.

\bibitem{barbadillo_2019}
S.~P. Guillermo~Barbadillo, ``Veridas solution for sdsv challenge technical
  report,'' 2019.

\bibitem{9003979}
D.~{Raj}, D.~{Snyder}, D.~{Povey}, and S.~{Khudanpur}, ``Probing the
  information encoded in x-vectors,'' in \emph{2019 IEEE Automatic Speech
  Recognition and Understanding Workshop (ASRU)}, 2019, pp. 726--733.

\bibitem{wang2017does}
S.~Wang, Y.~Qian, and K.~Yu, ``What does the speaker embedding encode?'' in
  \emph{Interspeech}, 2017, pp. 1497--1501.

\bibitem{zeinali2019but}
H.~Zeinali, S.~Wang, A.~Silnova, P.~Mat{\v{e}}jka, and O.~Plchot, ``But system
  description to voxceleb speaker recognition challenge 2019,'' 2019.

\bibitem{snyder2015musan}
D.~Snyder, G.~Chen, and D.~Povey, ``Musan: A music, speech, and noise corpus,''
  2015.

\bibitem{chung2018voxceleb2}
J.~S. Chung, A.~Nagrani, and A.~Zisserman, ``Voxceleb2: Deep speaker
  recognition,'' 2018.

\bibitem{mclaren2016speakers}
M.~McLaren, L.~Ferrer, D.~Castan, and A.~Lawson, ``The speakers in the wild
  (sitw) speaker recognition database.'' in \emph{Interspeech}, 2016, pp.
  818--822.

\bibitem{nagrani2017voxceleb}
A.~Nagrani, J.~S. Chung, and A.~Zisserman, ``Voxceleb: a large-scale speaker
  identification dataset,'' \emph{Interspeech}, pp. 2616--2620, 2017.

\bibitem{panayotov2015librispeech}
V.~Panayotov, G.~Chen, D.~Povey, and S.~Khudanpur, ``Librispeech: an asr corpus
  based on public domain audio books,'' in \emph{IEEE international conference
  on acoustics, speech and signal processing (ICASSP)}.\hskip 1em plus 0.5em
  minus 0.4em\relax IEEE, 2015, pp. 5206--5210.

\end{thebibliography}

\end{document}